\documentstyle[12pt,epsfig]{article}

\newcommand{\noi}{\noindent}
\newcommand{\beq}[1]{\begin{equation}\label{#1}}
\newcommand{\eeq}{\end{equation}}
\newcommand{\bear}[1]{\begin{eqnarray}\label{#1}}
\newcommand{\ear}{\end{eqnarray}}
\newcommand{\nn}{\nonumber}

\catcode`\@=11 \@addtoreset{equation}{section}\catcode`\@=12

\newcommand{\np}{ {\newpage } }
\newcommand{\N}{ \mbox{\rm I$\!$N} }
\newcommand{\R}{ \mbox{\rm I$\!$R} }
\def\C{\mbox{\rm {I\kern-.520em C}}}

\newcommand{\sign}{ \mbox{\rm sign} }

\newcommand{\e}{ \mbox{\rm e} }
\newcommand{\eps}{ \varepsilon }
\newcommand{\p}{\partial}
\newcommand{\btd}{\bigtriangledown}
\newcommand{\btu}{\bigtriangleup}
\newcommand{\tri}{\Delta}

\newcommand{\sums}{\sum\limits}
\newcommand{\const}{\mathop{\rm const}\nolimits}

\begin{document}

\thispagestyle{empty}

\begin{center}{ \large \bf
Billiard Representation for
Multidimensional Cosmology
\\ with Intersecting $p$-branes
near the Singularity}

\end{center}

\bigskip

\centerline{\bf \large
V. D. Ivashchuk and V. N. Melnikov}

\vspace{0.96truecm}

\centerline{Center for Gravitation and Fundamental Metrology}
\centerline{VNIIMS, 3-1 M. Ulyanovoy Str.}
\centerline{Moscow, 117313, Russia}
\centerline{e-mail: ivas@rgs.phys.msu.su}

\begin{abstract}
Multidimensional model describing the cosmological evolution
of  $n$
Einstein spaces in the theory with $l$ scalar fields and forms is
considered.  When electro-magnetic composite
$p$-brane ansatz is adopted,
and certain restrictions on the parameters of the model are
imposed, the dynamics of the model near the singularity is reduced
to a billiard on the $(N-1)$-dimensional Lobachevsky space $H^{N-1}$,
$N = n+l$.
The geometrical criterion for the finiteness of the billiard volume
and its compactness is used. This criterion reduces the problem
to the problem of illumination of $(N-2)$-dimensional sphere $S^{N-2}$
by point-like sources. Some examples with
billiards of finite volume and hence oscillating
behaviour near the singularity are considered.
Among them  examples with  square and triangle
2-dimensional billiards (e.g. that of the Bianchi-IX model)
and a 4-dimensional billiard
in ``truncated'' $D = 11$ supergravity model (without the Chern-Simons
term) are considered. It is shown that the inclusion of the Chern-Simons
term destroys  the confining of a billiard.

\end{abstract}

\bigskip

\hspace*{0.950cm} PACS number(s):\ 04.50.+h,\ 98.80.Hw,\ 04.60.Kz
\np

\section{Introduction}
\setcounter{equation}{0}

At present there exists a special interest to the so-called M-
and F-theories  \cite{HTW,Sc,Du,Vafa}.
These theories are ``supermembrane'' analogues of
superstring models \cite{GrSW} in $D=11,12$ etc. The low-energy limit of
these theories leads to models governed by the action
\beq{2.1i}
S = \int_{M} d^{D}z \sqrt{|g|} \{ {R}[g]  - 2\Lambda
- h_{\alpha\beta}\;
g^{MN} \partial_{M} \varphi^\alpha \partial_{N} \varphi^\beta
- \sum_{a \in \Delta}
\frac{\theta_a}{n_a!} \exp[ 2 \lambda_{a} (\varphi) ] (F^a)^2_g \},
\eeq
where $g = g_{MN} dz^{M} \otimes dz^{N}$ is a metric,
$\varphi=(\varphi^\alpha)\in \R^l$
is a vector from dilatonic scalar fields,
$(h_{\alpha\beta})$ is a positively-defined  symmetric
$l\times l$ matrix ($l\in \N$), $\theta_a = \pm 1$,
\beq{2.2i}
F^a =  dA^a =
\frac{1}{n_a!} F^a_{M_1 \ldots M_{n_a}}
dz^{M_1} \wedge \ldots \wedge dz^{M_{n_a}}
\eeq
is a $n_a$-form ($n_a \geq 1$) on a $D$-dimensional manifold $M$,
$D > 2$,  $\Lambda$ is cosmological constant and $\lambda_{a}$ is a
$1$-form on $\R^l$: $\lambda_{a} (\varphi) =\lambda_{a \alpha}
\varphi^\alpha$, $a \in \Delta$, $\alpha=1,\ldots,l$. In (\ref{2.1i}) we
denote $|g| = |\det (g_{MN})|$, \beq{2.3i} (F^a)^2_g = F^a_{M_1 \ldots
M_{n_a}} F^a_{N_1 \ldots N_{n_a}} g^{M_1 N_1} \ldots g^{M_{n_a} N_{n_a}},
\eeq
$a \in \Delta$, where $\Delta$ is some finite (non-empty) set. In
models with one time all $\theta_a =  1$  when the signature of the metric
is $(-1,+1, \ldots, +1)$.

In \cite{IMC} it was shown that after dimensional reduction on the
manifold
\beq{1.1}
M_{*}\times M_1\times\dots\times M_n
\eeq
with $M_i$ being Einstein space ($i = 1, \ldots, n$)
and when the composite
$p$-brane ansatz is considered
(for review see, for example, \cite{IMC,DKL,St,BREJS,AV,AR})
the problem is reduced to the gravitating
self-interacting $\sigma$-model with certain constraints imposed. For
electric $p$-branes see also \cite{IM0,IMO,IMR}
(in  \cite{IMR} the composite electric case was considered).

In cosmological (or spherically symmetric) case $M_{*} = \R$ and
the problem is effectively reduced to a
Toda-like system with the Lagrangian \cite{IMJ}
\beq{1.2}
L=\frac12 \bar G_{AB}\dot x^A\dot x^B-
\sum_{s\in S_{*}} A_s \exp(2U^s_A x^A),
\eeq
and the zero-energy constraint $E = 0$ imposed, where
$(\bar G_{AB})$  is a non-degenerate symmetric
$N \times N$ matrix ($N = n + l$),
$A_s \neq 0$, $x = (x^A) \in \R^N$, $U^s = (U^s_A) \in \R^N$,
$s \in S_{*}$.
The considered cosmological model contains some stringy cosmological
models (see for example \cite{LMPX}). It may be obtained (at a classical
level) from a multidimensional cosmological model with a perfect fluid
\cite{IM3,GIM} as a special case.

The integrability of the Lagrange equations corresponding to
(\ref{1.2}) crucially depends upon the scalar
products $(U^{s_1},U^{s_2})$, $s_1, s_2 \in S_{*}$,
where
\beq{1.3}
(U,U')=\bar G^{AB} U_A U'_B,
\eeq
$U,U'\in \R^N$, where $(\bar G^{AB})=(\bar G_{AB})^{-1}$.

In the orthogonal case
\beq{1.4}
(U^s,U^{s'})=0
\eeq
$s, s' \in S_{*}$,
a class of cosmological and spherically-symmetric solutions was obtained
in \cite{IMJ}. Special cases were also considered in
\cite{LPX,BGIM,BIM,GrIM}. Recently the ``orthogonal''
solutions were generalized to so-called  ``block orthogonal''
case \cite{Br1,IMJ1}.

This paper is devoted to the investigation of the possible
oscillating (and probably stochastic) behaviour near the
singularity (see \cite{Mi}-\cite{IM} and references therein)
for cosmological models with $p$-branes.

We remind that
near the singularity one can have an oscillating behavior like in
the well-known mixmaster  (Bianchi-IX) model \cite{Mi}-\cite{MTW}
(see also \cite{P}-\cite{M1}).
Multidimensional generalizations and analogues of this model
were considered by many authors (see, for
example, \cite{BK}-\cite{CDDQ}). In \cite{IKM1,IKM,IM} a billiard
representation for a multidimensional cosmological models near the
singularity was considered and a criterion for a volume of the
billiard to be finite was established in terms of illumination of the unit
sphere by point-like sources. For perfect-fluid this was considered in
detail in \cite{IM}. Some topics related to general
(non-homogeneous) situation were considered in \cite{KM}.

Here we apply the billiard approach
suggested in \cite{IKM1,IKM,IM} to a
$p$-brane cosmology. The cosmological model with $p$-branes
may be considered as a special case of a  cosmological
model for multicomponent perfect fluid
with the equations of state for ``brane'' components:
$p_i^s = - \rho^s$ or
$p_i^s =  \rho^s$, when brane $s$ ``lives'' or ``does not live''
in the space $M_i$ respectively, $i = 1, \ldots, n$.

The paper is organized as follows. In Sec. 2 the cosmological
model with $p$-branes is considered. Sec. 3 deals with the
Lagrange representation to equations of motion and the diagonalization
of the Lagrangian. In Sec. 4 a billiard approach in the
multidimensional cosmology with $p$-branes is developed.
A necessary condition for
the existence of oscillating (e.g. stochastic behaviour)
near the singularity is established:
\beq{1.5}
m  \geq n+l,
\eeq
where $m$ is  the number of $p$-branes.
In Sections 5 and 6 some examples of billiards
(e.g. in  truncated $D=11$ supergravity etc.) are considered.

\section{\bf The model}
\setcounter{equation}{0}

Equations of motion corresponding to  (\ref{2.1i}) have the following
form
\bear{2.4i}
R_{MN} - \frac{1}{2} g_{MN} R  =   T_{MN},
\\
\label{2.5i}
{\btu}[g] \varphi^\alpha -
\sum_{a \in \Delta} \theta_a  \frac{\lambda^{\alpha}_a}{n_a!}
e^{2 \lambda_{a}(\varphi)} (F^a)^2_g = 0,
\\
\label{2.6i}
\nabla_{M_1}[g] (e^{2 \lambda_{a}(\varphi)}
F^{a, M_1 \ldots M_{n_a}})  =  0,
\ear
$a \in \Delta$; $\alpha=1,\ldots,l$.
In (\ref{2.5i}) $\lambda^{\alpha}_{a} = h^{\alpha \beta}
\lambda_{a \beta }$, where $(h^{\alpha \beta})$
is a matrix inverse to $(h_{\alpha \beta})$.
In (\ref{2.4i})
\bear{2.7i}
T_{MN} =   T_{MN}[\varphi,g]
+ \sum_{a\in\Delta} \theta_a  e^{2 \lambda_{a}(\varphi)} T_{MN}[F^a,g],
\ear
where
\bear{2.8i}
T_{MN}[\varphi,g] =
h_{\alpha\beta}\left(\p_{M} \varphi^\alpha \p_{N} \varphi^\beta -
\frac{1}{2} g_{MN} \p_{P} \varphi^\alpha \p^{P} \varphi^\beta\right),
\\
T_{MN}[F^a,g] = \frac{1}{n_{a}!}\left[ - \frac{1}{2} g_{MN} (F^{a})^{2}_{g}
+ n_{a}  F^{a}_{M M_2 \ldots M_{n_a}} F_{N}^{a, M_2 \ldots M_{n_a}}\right].
\label{2.9i}
\ear
In (\ref{2.5i}), (\ref{2.6i}) ${\btu}[g]$ and ${\btd}[g]$
are the Laplace-Beltrami and covariant derivative operators respectively
corresponding to  $g$.

Let us consider a manifold
\beq{2.10g}
M = \R  \times M_{1} \times \ldots \times M_{n}
\eeq
with the metric
\beq{2.11g}
g= w \e^{2{\gamma}(t)} dt \otimes dt +
\sum_{i=1}^{n} \e^{2\phi^i(t)} g^i ,
\eeq
where $w=\pm 1$, $t$ is a distinguished coordinate which, by
convention, will be called ``time'';
$g^i  = g^i_{m_{i} n_{i}}(y_i) dy_i^{m_{i}} \otimes dy_i^{n_{i}}$
is a metric on $M_{i}$  satisfying the equation
\beq{2.12g}
R_{m_{i}n_{i}}[g^i ] = \xi_{i} g^i_{m_{i}n_{i}},
\eeq
$m_{i},n_{i}=1,\ldots,d_{i}$; $d_{i} = \dim M_i$, $\xi_i= \const$,
$i=1,\dots,n$; $n \in \N$. Thus, $(M_i,g^i)$ are Einstein spaces.
The functions $\gamma,\phi^i$: $(t_-,t_+)\to\R$ are smooth.

{\bf Remark.}
{\em It is more correct to write in (\ref{2.11g})
$\hat{g}^{i}$ instead of $g^{i}$, where $\hat{g}^{i} = p_{i}^{*} g^{i}$
is the pullback of the metric $g^{i}$  to the manifold  $M$ by the
canonical projection: $p_{i} : M \rightarrow  M_{i}$, $i = 1,
\ldots, n$. In what follows we omit "hats" for  simplicity.}

Each manifold $M_i$ is assumed to be oriented and connected,
$i = 1,\ldots,n$. Then the volume $d_i$-form
\beq{2.13g}
\tau_i  = \sqrt{|g^i(y_i)|}
\ dy_i^{1} \wedge \ldots \wedge dy_i^{d_i},
\eeq
and the signature parameter
\beq{2.14g}
\eps(i)  = \sign \det (g^i_{m_{i}n_{i}}) = \pm 1
\eeq
are correctly defined for all $i=1,\ldots,n$.

Let
\beq{2.15g}
\Omega_0 = \{ \emptyset, \{ 1 \}, \ldots, \{ n \},
\{ 1, 2 \}, \ldots, \{ 1,  \ldots, n \} \}
\eeq
be a set of all subsets of
\beq{2.25n}
I_0\equiv\{ 1, \ldots, n \}.
\eeq
Let $I = \{ i_1, \ldots, i_k \} \in \Omega_0$, $i_1 < \ldots < i_k$.
We define a form
\beq{2.17i}
\tau(I) \equiv \tau_{i_1}  \wedge \ldots \wedge \tau_{i_k},
\eeq
of rank
\beq{2.19i}
d(I) \equiv  \sum_{i \in I} d_i,
\eeq
and a corresponding $p$-brane submanifold
\beq{2.18i}
M_{I} \equiv M_{i_1}  \times  \ldots \times M_{i_k},
\eeq
where $p=d(I)-1$ (${\rm dim M_{I}} = d(I)$).
We also define $\eps$-symbol
\beq{2.19e}
\eps(I) \equiv  \eps(i_1) \ldots \eps(i_k).
\eeq
For $I = \emptyset$ we put  $\tau(\emptyset) = \eps(\emptyset) = 1$,
 $d(\emptyset) = 0$.

For fields of forms we adopt the following "composite electro-magnetic"
ansatz
\beq{2.27n}
F^a=
\sum_{I\in\Omega_{a,e}}{\cal F}^{(a,e,I)}+\sum_{J\in\Omega_{a,m}}{\cal F}^{(a,m,J)},
\eeq
where
\bear{2.28n}
{\cal F}^{(a,e,I)}=d\Phi^{(a,e,I)}\wedge\tau(I), \\ \label{2.29n}
{\cal F}^{(a,m,J)}=\e^{-2\lambda_a(\varphi)}
*\left(d\Phi^{(a,m,J)}\wedge\tau(J)\right),
\ear
$a\in\tri$, $I\in\Omega_{a,e}$, $J\in\Omega_{a,m}$ and
\beq{2.29nn}
\Omega_{a,e},\Omega_{a,m}\subset \Omega_0.
\eeq
(For empty $\Omega_{a,v}=\emptyset$, $v=e,m$, we put $\sums_\emptyset=0$ in
(\ref{2.27n})). In (\ref{2.29n}) $*=*[g]$ is the Hodge operator on $(M,g)$.

For  potentials in (\ref{2.28n}), (\ref{2.29n}) we put
\beq{2.28nn}
\Phi^s=\Phi^s(t),
\eeq
$s\in S$, where
\beq{6.39i}
S=S_e \sqcup S_m,  \qquad
S_v \equiv \sqcup_{a\in\tri}\{a\}\times\{v\}\times\Omega_{a,v},
\eeq
$v=e,m$. Here $\sqcup$ means the union of non-intersecting sets.
The set  $S$ consists of elements
$s=(a_s,v_s,I_s)$, where $a_s \in \tri$, $v_s = e,m$ and
$I_s \in \Omega_{a,v_s}$ are ``color'', ``electro-magnetic'' and
``brane'' indices, respectively.

For dilatonic scalar fields we put
\beq{2.30n}
\varphi^\alpha=\varphi^\alpha(t),
\eeq
$\alpha=1,\dots,l$.

From  (\ref{2.28n})  and (\ref{2.29n}) we obtain
the relations between dimensions of $p$-brane
worldsheets and ranks of forms
\bear{2.d1}
d(I) = n_a - 1,  \quad I \in \Omega_{a,e},
\\ \label{2.d2}
d(J) = D - n_a - 1,  \quad J \in \Omega_{a,m},
\ear
in electric and magnetic cases respectively.

\section{\bf Lagrange representation}
\setcounter{equation}{0}

Here, like in \cite{IMJ}, we impose a restriction on
$p$-brane configurations, or, equivalently, on $\Omega_{a,v}$.
We assume that the energy momentum tensor $(T_{MN})$ has a block-diagonal
structure (as it takes place for $(g_{MN})$).
Sufficient  restrictions on $\Omega_{a,v}$ that guarantee
a block-diagonality of $(T_{MN})$ are the following ones:

{\bf 1.} for any $a \in \tri$ and $v=e,m$ there are no
$I, J \in \Omega_{a,v}$ such that
\beq{B.2}
I= \{i\}\sqcup(I\cap J), \  J=\{j\}\sqcup(I\cap J),
\quad d_i = d_j = 1, \ i \neq j;
\eeq

{\bf 2.} for any $a \in \tri$  there are no
$I \in \Omega_{a,m}$ and $J \in \Omega_{a,e}$
such that
\beq{B.1}
\bar I=\{j\}\sqcup J, \quad d_j = 1,
\eeq
$i,j=1, \ldots, n$.

In (\ref{B.1})
\beq{B.28i}
\bar I \equiv I_0 \setminus I
\eeq
is a  ``dual'' set ($I_0$ is defined in (\ref{2.25n})).
The restrictions (\ref{B.2}) and (\ref{B.1}) are
trivially satisfied when $n_1\le1$ and $n_1=0$ respectively, where
$n_1$ is the number of $1$-dimensional manifolds among $M_i$.

It follows from \cite{IMC} (see Proposition 2 in \cite{IMC}) that the
equations of motion (\ref{2.4i})--(\ref{2.6i}) and the Bianchi
identities
\beq{2.b}
d{\cal F}^s=0, \quad s\in S,
\eeq
for the field configuration (\ref{2.11g}), (\ref{2.27n})--(\ref{2.29n}),
(\ref{2.28nn}), (\ref{2.30n}) with the restrictions (\ref{B.2}),
(\ref{B.1}) imposed are equivalent to equations of motion for
Lagrange system with the Lagrangian
\bear{2.25gn}
L = \frac{1}{2}
{\cal N}^{-1} \biggl\{G_{ij}\dot\phi^i\dot\phi^j
+h_{\alpha\beta}\dot\varphi^{\alpha}\dot\varphi^{\beta}
+ \sum_{s\in S}\eps_s\exp[-2U^s(\phi,\varphi)](\dot\Phi^s)^2
-2{\cal N}^{2}V(\phi)\biggr\},
\ear
where $\dot x\equiv dx/dt$,
\beq{2.27gn}
V = {V}(\phi) =
(-w\Lambda)\e^{2{\gamma_0}(\phi)} +
\frac w2\sum_{i =1}^{n} \xi_i d_i
\e^{-2 \phi^i + 2 {\gamma_0}(\phi)}
\eeq
is a potential with
\beq{2.24gn}
\gamma_0 = \gamma_0(\phi)
\equiv\sum_{i=1}^nd_i\phi^i,  \label{2.32g}
\eeq
and
\beq{2.24gn1}
{\cal N}=\exp(\gamma-\gamma_0)>0
\eeq
is the lapse function,
\bear{2.u}
 U^s = U^s(\phi,\varphi)= -\chi_s\lambda_{a_s}(\varphi) +
\sum_{i\in I_s}d_i\phi^i, \\ \label{2.e}
\eps_s=(-\eps[g])^{(1-\chi_s)/2}\eps(I_s)\theta_{a_s}
\ear
for $s=(a_s,v_s,I_s)\in S$, $\eps[g]= \sign \det (g_{MN})$,
(more explicitly, (\ref{2.e}) reads: $\eps_s=\eps(I_s) \theta_{a_s}$ for
$v_s = e$ and $\eps_s=-\eps[g] \eps(I_s) \theta_{a_s}$ for
$v_s = m$)
\bear{2.x1}
\chi_s=+1, \quad v_s=e; \\ \label{2.x2}
\chi_s=-1, \quad v_s=m,
\ear
and
\beq{2.c}
G_{ij}=d_i\delta_{ij}-d_id_j
\eeq
are components of a ``pure cosmological'' minisupermetric;
$i,j=1,\dots,n$ \cite{IMZ}.

Let $x=(x^A)=(\phi^i,\varphi^\alpha)$,
\bear{2.35n}
\bar G=\bar G_{AB}dx^A\otimes dx^B=G_{ij}d\phi^i\otimes d\phi^j+
h_{\alpha\beta}d\varphi^\alpha\otimes d\varphi^\beta, \\ \label{2.36n}
(\bar G_{AB})=\left(\begin{array}{cc}
G_{ij}&0\\
0&h_{\alpha\beta}
\end{array}\right),
\ear
$U^s(x)=U_A^sx^A$ is defined in  (\ref{2.u}) and
\beq{2.38n}
(U_A^s)=(d_i\delta_{iI_s},-\chi_s\lambda_{a_s\alpha}).
\eeq
Here
\beq{2.39n}
\delta_{iI}\equiv\sum_{j\in I}\delta_{ij}=\begin{array}{ll}
1,&i\in I\\
0,&i\notin I
\end{array}
\eeq
is an indicator of $i$ belonging to $I$. The potential (\ref{2.27gn})
reads
\beq{2.40n}
V= (-w\Lambda)\e^{2U^\Lambda(x)} +
\sum_{j=1}^n\frac w2\xi_jd_j \e^{2U^j(x)},
\eeq
where
\bear{2.41n}
U^j(x)=U_A^jx^A=-\phi^j+\gamma_0(\phi),
\\ \label{2.42n}
U^\Lambda(x)=U_A^\Lambda x^A=\gamma_0(\phi),
\\ \label{2.43n}
(U_A^j)=(-\delta_i^j+d_i,0)
\\ \label{2.44n}
(U_A^\Lambda)=(d_i,0).
\ear

The integrability of the Lagrange system (\ref{2.25gn}) depends
upon the scalar products of the co-vectors  $U^{\Lambda}$, $U^j$, $U^s$
corresponding to $\bar G$:
\beq{2.45n}
(U,U')=\bar G^{AB}U_AU'_B,
\eeq
where
\beq{2.46n}
(\bar G^{AB})=\left(\begin{array}{cc}
G^{ij}&0\\
0&h^{\alpha\beta}
\end{array}\right)
\eeq
is a matrix inverse to (\ref{2.36n}). Here (as in \cite{IMZ})
\beq{2.47n}
G^{ij}=\frac{\delta^{ij}}{d_i} + \frac1{2-D},
\eeq
$i,j=1,\dots,n$. These products have the following form
\cite{IMC}
\bear{2.48n}
(U^i,U^j)=\frac{\delta_{ij}}{d_j}-1,
\\ \label{2.50n}
(U^\Lambda,U^\Lambda)=-\frac{D-1}{D-2},
\\ \label{2.51n}
(U^s,U^{s'})=
d(I_s \cap I_{s'})+\frac{d(I_s)d(I_{s'})}{2-D}
+\chi_s \chi_{s'} \lambda_{a\alpha}\lambda_{b\beta}
h^{\alpha\beta},
\\ \label{2.52n}
(U^s,U^i)=-\delta_{iI_s},
\\ \label{2.53n}
(U^{\Lambda}, U^i)=- 1,
\\ \label{2.54n}
(U^{\Lambda}, U^s)= \frac{d(I_s)}{2-D},
\ear
where $s=(a_s,v_s,I_s)$, $s'=(a_{s'},v_{s'},I_{s'})\in S$.

First we integrate the ``Maxwell equations'' (for $s\in S_e$)
and Bianchi identities (for $s\in S_m$):
\bear{5.29n}
\frac d{dt}\left(\exp(-2U^s)\dot\Phi^s\right)=0
\Longleftrightarrow
\dot\Phi^s=Q_s \exp(2U^s),
\ear
where $Q_s$ are constants, $s \in S$.  We put
\bear{5.30n}
Q_s \ne 0,
\ear
for all $s \in S$.

For fixed charges
$Q=(Q_s, s \in S)$  Lagrange equations for the Lagrangian
(\ref{2.25gn}) corresponding to $(x^A)=(\phi^i,\varphi^\alpha)$,
(when relations   (\ref{5.29n})
are substituted) are equivalent to
Lagrange equations for the Lagrangian \cite{IMJ}
\beq{5.31n}
L_Q=\frac12 {\cal N}^{-1} \bar G_{AB} \dot x^A\dot x^B-
{\cal N} V_Q.
\eeq
where
\beq{5.32n}
V_Q=V+\frac12\sum_{s\in S}\eps_sQ_s^2\exp[2U^s(x)],
\eeq
$(\bar G_{AB})$ and $V$ are defined in (\ref{2.36n}) and (\ref{2.40n})
respectively.

\subsection{ Diagonalization of the Lagrangian}

The minisuperspace metric (\ref{2.35n})
has a pseudo-Euclidean signature $(-,+, \ldots ,+)$,
since the matrix $(G_{ij})$ has the pseudo-Euclidean
signature, and $(h_{\alpha \beta})$ has the Euclidean one.
Hence there exists
a linear transformation
\beq{2.21o}
z^{a}=e^{a}_{A}x^{A},
\eeq
diagonalizing the minisuperspace metric (\ref{2.35n})
\beq{2.22o}
\bar G= \eta_{ab}dz^{a} \otimes dz^{b}=
 -dz^{0} \otimes dz^{0} + \sum_{k=1}^{N-1}dz^{k}\otimes dz^{k},
\eeq
where
\beq{2.23o}
(\eta_{ab})=(\eta^{ab}) \equiv diag(-1,+1, \ldots ,+1),
\eeq
and here and in what follows
$a,b = 0, \ldots ,N-1$; $N =n+l$.
The matrix of  linear transformation
$(e^{a}_{A})$  satisfies the relation
\beq{2.24o}
\eta_{ab} e^{a}_{A} e^{b}_{B} = \bar{G}_{AB}
\eeq
or, equivalently,
\beq{2.25o}
\eta^{ab} = e^{a}_{A}\bar{G}^{AB} e^{b}_{B} =  (e^{a},e^{b}),
\eeq
where $e^a = (e^a_A)$.

Inverting the map (\ref{2.21o}) we get
\beq{2.28o}
x^{A} = e_{a}^{A} z^{a},
\eeq
where for components of the inverse matrix
$(e_{a}^{A}) = (e^{a}_{A})^{-1}$ we obtain from (\ref{2.25o})
\beq{2.29o}
e_{a}^{A}    = \bar{G}^{AB} e^{b}_{B} \eta_{ba}.
\eeq

Like in \cite{IM} we put
\beq{2.30o}
e^{0} = q^{-1} U^{\Lambda}, \qquad
q = [(D-1)/(D-2)]^{1/2}= [- (U^\Lambda,U^\Lambda)]^{1/2}.
\eeq
and hence
\beq{2.31o}
z^0 = e^{0}_{A} x^A = \sum_{i=1}^{n} q^{-1} d_i x^i.
\eeq

In $z$-coordinates (\ref{2.21o}) with $z^0$ from
(\ref{2.31o}) the Lagrangian (\ref{5.31n}) reads
\beq{2.32o}
L_Q = {L}_Q(z^{a}, \dot{z}^{a}, {\cal N})
= \frac{1}{2} {\cal N}^{-1}
\eta_{ab} \dot{z}^{a} \dot{z}^{b} -  {\cal N} {V}(z),
\eeq
where
\beq{2.34o}
{V}(z) =
\sum_{r \in S_{*}} A_{r} \exp(2 U^{r}_a z^a)
\eeq
is a potential,
\beq{2.34oa}
S_{*} = \{ \Lambda \} \cup \{ 1, \ldots, n \} \cup S
\eeq
is an index set and
\beq{2.34ob}
A_{\Lambda} = -w \Lambda, \quad A_j = \frac{w}{2} \xi_j d_j, \quad
A_s = \frac{1}{2}\eps_s Q_s^2,
\eeq
$j = 1, \ldots, n$; $s \in S$.
Here we denote
\beq{2.35o}
U^{r}_a  = e_{a}^{A} U^{r}_A =
(U^{r}, e^{b}) \eta_{ba},
\eeq
$a = 0, \ldots , N-1$; $r \in S_{*}$ (see (\ref{2.29o})).

From (\ref{2.45n})-(\ref{2.47n}), (\ref{2.30o}) and (\ref{2.35o})
we deduce
\beq{2.36o}
U^{r}_0  = - (U^{r}, e^{0}) =
(\sum_{i=1}^{n} U^{r}_i ) / q(D-2),
\eeq
$r \in S_*$.
For the $\Lambda$-term and curvature-term components
we obtain from  (\ref{2.30o})  and (\ref{2.36o})
\beq{2.37o}
U^{\Lambda}_0 = q > 0 , \qquad U^{j}_0 = 1/q > 0,
\eeq
$j= 1,\ldots,n$.

For ``brane'' components we get
from (\ref{2.38n}), (\ref{2.30o})  and  (\ref{2.36o})
\beq{2.38o}
U^{s}_0 = d(I_s)/\sqrt{(D-2)(D-1)} > 0.
\eeq

We remind that  (see (\ref{2.48n}) and  (\ref{2.50n}))
that
\beq{2.39o}
(U^{\Lambda},U^{\Lambda}) = (D-1)/(2-D) < 0, \qquad  (U^{j},U^{j}) =
(\frac{1}{d_j} - 1) < 0,
\end{equation}
for $d_j > 1$, $j= 1,\ldots,n$. For $d_j = 1$ we have $\xi^{j}
= A_{j} = 0$.

\section{Billiard representation}
\setcounter{equation}{0}

Here we put the following restrictions on parameters of the model:
\bear{4.1n}
{\bf (i)}  \qquad \qquad \eps_s = + 1,
\\ \label{4.2n}
{\bf (ii)} \quad  d(I_s) < D-2,
\ear
$s \in S$.
For $\theta_a = 1$, $a \in \Delta$, and $\eps[g] = -1$,
the first restriction means that all $\eps(I_s) = 1$, $s \in S$, i.e.
all $p$-branes are either Euclidean or contain even number of
``times''. Restriction ${\bf (ii)}$  implies
\beq{4.3}
(U^s,U^{s})=
d(I_s ) \left(1 +\frac{d(I_s)}{2-D} \right)
+ \lambda_{a_s}^2 >0,
\eeq
where $\lambda^2 = \lambda_{\alpha} \lambda_{\beta}
h^{\alpha\beta}$, $D>2$. As we shall see below, both restrictions
are necessary for a formation of potential walls
in the Lobachevsky space when a certain asymptotic in time variable
is considered.

Here we consider a behaviour of the dynamical system, described
by the Lagrangian (\ref{2.32o}) with the potential (\ref{2.34o})
for $N \geq 3$ in the limit
\beq{3.1o}
z^0  \rightarrow  -\infty, \qquad z =(z^0, \vec{z}) \in {\cal V}_{-},
\eeq
where  ${\cal V}_{-}
\equiv \{(z^0, \vec{z}) \in \R^N | z^0 < - |\vec{z}| \}$
is the lower light cone. For the volume scale factor
\beq{3.2o}
v = \exp(\sum_{i=1}^{n} d_{i}x^{i}) = \exp(qz^0)
\eeq
($q > 0$) we have in this limit $v \rightarrow + 0$.
Under certain additional assumptions the limit (\ref{3.1o}) describes
an approaching to the singularity.

Due to relations
(\ref{2.34ob}), (\ref{2.37o})-(\ref{2.39o}), (\ref{4.1n}) and (\ref{4.3})
the parameters $U^r$  in the potential (\ref{2.34o})
obey the following restrictions:
\bear{3.3o}
&& {\bf 1.} \ A_r > 0  \
{\rm for} \ (U^r)^2 = -(U^r_0)^2 + (\vec{U}^r)^2 > 0;
\\  \label{3.4o}
&& {\bf 2.} \ U^r_0 > 0  \ {\rm for \ all} \  r \in S_{*}.
\ear

Now we restrict the Lagrange system (\ref{2.32o}) on ${\cal V}_{-}$, i.e.
we consider the Lagrangian
\beq{3.5o}
L_{-} \equiv L|_{TM_{-}} , \qquad M_{-} = {\cal V}_{-} \times \R_{+},
\end{equation}
where $ TM_{-}$ is a tangent vector bundle over $M_{-}$  and
$\R_{+} \equiv \{ {\cal N} > 0 \}$. (Here $F|_{A}$ means a restriction
of function $F$ on $A$.)
Introducing  an analogue of the Misner-Chitre coordinates in
$\cal{V}_{-}$ [36,37] which reduce the problem to a unit disk
\bear{3.6o}
&&z^0 = - \exp(-y^0) \frac{1 + \vec{y}^2}{1 - \vec{y}^2}, \\
&&\vec{z} = - 2 \exp(-y^0) \frac{ \vec{y}}{1 - \vec{y}^2},
\ear
$|\vec{y}| < 1$, we get for the Lagrangian (\ref{2.32o})
\beq{3.8o}
L_{-} = \frac{1}{2} {\cal N}^{-1} e^{- 2 y^0}
[- (\dot{y}^{0})^2 + \bar{h}_{ij}(\vec{y}) \dot{y}^{i} \dot{y}^{j}]
-  {\cal N} V.
\eeq
Here
\beq{3.9o}
\bar{h}_{ij}(\vec{y}) = 4 \delta_{ij} (1 - \vec{y}^2)^{-2},
\eeq
$i,j =1, \ldots , N-1$, and
\beq{3.10o}
V = {V}(y) =
\sum_{r \in S_*} A_r \exp[\bar{\Phi}(y,2U^r)],
\end{equation}
where
\beq{3.11o}
{\bar{\Phi}}(y,u)  \equiv - e^{-y^0}(1 - \vec{y}^2)^{-1}
[u_0 (1 + \vec{y}^2) + 2 \vec{u}\vec{y}],
\eeq

We note that the $(N-1)$-dimensional open unit disk (ball)
\beq{3.12o}
D^{N-1} \equiv \{ \vec{y}= (y^1, \ldots, y^{N - 1})| |\vec{y}| < 1 \}
\subset \R^{N-1}
\eeq
with the metric $\bar{h} = \bar{h}_{ij}(\vec{y})
dy^i \otimes dy^j $ is one
of the realization of the $(N-1)$-dimensional Lobachevsky space
$H^{N-1}$.

We fix the gauge
\beq{3.13o}
{\cal N} =   \exp(- 2y^0) = - z^2.
\eeq
Then, it is not difficult to verify that the
Lagrange equations for the Lagrangian (\ref{3.8o}) with the gauge fixing
(\ref{3.3o})  are  equivalent  to  the Lagrange equations for the
Lagrangian
\beq{3.14o}
L_{*} = - \frac{1}{2}  (\dot{y}^{0})^2 +  \frac{1}{2}
\bar{h}_{ij}(\vec{y}) \dot{y}^{i} \dot{y}^{j} -  V_{*}
\eeq
with the energy constraint imposed
\beq{3.15o}
E_{*}
= - \frac{1}{2}  (\dot{y}^{0})^2 +  \frac{1}{2}
\bar{h}_{ij}(\vec{y}) \dot{y}^{i} \dot{y}^{j} +  V_{*} = 0.
\eeq
Here
\beq{3.16o}
V_{*} =  e^{-2y^0} V =
\sum_{r \in S_*} A_{\alpha} \exp[\Phi(y,2U^r)],
\eeq
where
\beq{3.17o}
{\Phi}(y,u)  = - 2y^0 + \bar{\Phi}(y,u).
\eeq

Now we are interested in a behavior of the dynamical system
in the limit  $y^0 \rightarrow - \infty$ (or, equivalently, in
the limit $z^2  = -(z^0)^2 + (\vec{z})^2
\rightarrow - \infty$, $z^0 < 0$) implying (\ref{3.1o}).
Using the relations  ($U_0 \neq 0$ )
\bear{3.18o}
&&{\Phi}(y,2U) = - 2 U_0 \exp(-y^0)
\frac{{A}(\vec{y}, -\vec{U}/U_0)}{1 - \vec{y}^2} - 2y^0, \\
&& {A}(\vec{y}, \vec{v}) \equiv
(\vec{y} - \vec{v})^2 -\vec{v}^2 + 1,
\ear
we get
\beq{3.20o}
\lim_{y^0 \rightarrow - \infty} \exp {\Phi}(y, 2U)  = 0
\eeq
for $U^2 = - (U_0)^2 + (\vec{U})^2 \leq 0$, $U_0 > 0$ and
\beq{3.21o}
\lim_{y^0 \rightarrow - \infty} \exp {\Phi}(y,U)  =
{\theta_{\infty}}(-{A}(\vec{y}, - \vec{U}/U_0))
\eeq
for $U^2 > 0$, $U_0 > 0$. In (\ref{3.21o}) we denote
\bear{3.22o}
{\theta_{\infty}}(x) \equiv + &\infty, &x \geq 0,  \nonumber \\
                              & 0    , &x < 0.
\ear
Using relations (\ref{3.3o}), (\ref{3.4o}) and relations (\ref{3.16o}),
(\ref{3.20o}), (\ref{3.21o}) we obtain
\beq{3.23o}
V_{\infty}(\vec{y}) \equiv
\lim_{y^0 \rightarrow - \infty} {V_{*}}(y^0, \vec{y}) = \sum_{s \in
S_{*}} {\theta_{\infty}}(-{A}(\vec{y}, -\vec{U^s}/U_0^s)).
\eeq

The potential $V_{\infty}$ may be  written as follows
\bear{3.25o}
{V_{\infty}}(\vec{y}) =
{V}(\vec{y},B) \equiv &0, &\vec{y} \in B,
\nonumber \\
&+ \infty, &\vec{y}
\in D^{N-1} \setminus B,
\ear
where
\beq{4.4}
B = \bigcap_{s \in S} {B}_{s}  \subset D^{N-1},
\eeq
\beq{4.5}
{B}_{s}  = \{ \vec{y} \in D^{N-1} |
|\vec{y} - \vec{v}^{s}| > r_{s} \},
\eeq
where
\beq{4.6}
\vec{v}^{s} = - \vec{U}^s/U^s_0,
\eeq
($|\vec{v^s}| > 1$) and
\beq{4.7}
r_s = \sqrt{(\vec{v}^s)^2 - 1},
\eeq
$s \in S$. Remind that  $(U^s_a) = (U^s_0, \vec{U}^s)$  is defined by
relation  (\ref{2.35o}).

$B$ is an open domain.
Its boundary $\partial B = \bar{B} \setminus B$ is formed by
certain parts of $m = |S|$ $(N-2)$-dimensional
spheres with the centers in the points $\vec{v}^s$
($|\vec{v^{\alpha}}| > 1$) and radii $r_s$, $s \in S$.

So, in the limit $y^{0} \rightarrow - \infty$ we are led to the
dynamical system
\bear{3.30o}
&L_{\infty} = - \frac{1}{2} (\dot{y}^{0})^2 +  \frac{1}{2}
\bar{h}_{ij}(\vec{y}) \dot{y}^{i} \dot{y}^{j} -  {V_{\infty}}(\vec{y}), \\
&E_{\infty} = - \frac{1}{2} (\dot{y}^{0})^2 +  \frac{1}{2}
\bar{h}_{ij}(\vec{y}) \dot{y}^{i} \dot{y}^{j} +  {V_{\infty}}(\vec{y}) = 0,
\ear
which after the separating of $y^0$ variable
\beq{3.32o}
y^0 = \omega (t - t_0),
\eeq
($\omega \neq 0$ , $t_0$  are constants) is reduced to the Lagrange
system with the Lagrangian
\beq{3.33o}
L_{B} =  \frac{1}{2} \bar{h}_{ij}(\vec{y})
\dot{y}^{i} \dot{y}^{j} -  {V}(\vec{y},B).
\eeq
Due to (\ref{3.32o})
\beq{3.34o}
E_{B} =  \frac{1}{2}
\bar{h}_{ij}(\vec{y}) \dot{y}^{i} \dot{y}^{j} +  {V}(\vec{y},B) =
\frac{\omega^2}{2}.
\eeq

We put $\omega > 0$, then the limit $t \rightarrow - \infty$ describes
an approach to the singularity.

When  $S \neq \emptyset$ the Lagrangian
(\ref{3.33o}) describes a motion of a particle  of  unit  mass,
moving in the ($N-1$)-dimensional billiard $B \subset D^{N-1}$  (see
(\ref{4.4})).  The geodesic motion in $B$
corresponds to a ``Kasner epoch'' and the reflection from the boundary
corresponds to the change of Kasner epochs \cite{IM}.

Let the billiard $B$  has an infinite volume: ${\rm vol} B = +\infty$ and
there are open zones  at  the  infinite  sphere
$|\vec{y}| =1$. After a finite number of reflections from the boundary
a  particle  moves  towards  one  of  these  open  zones.
In this case for a
corresponding cosmological model we get the ``Kasner-like''
behavior in the limit $t \rightarrow - \infty$ and the absence
of a stochastic behaviour.

Let ${\rm vol} B < + \infty$. There are two possibilities
in this case: i) the
closure of the billiard  $\bar{B}$        is
compact (in the topology of $D^{N-1}$) ; ii) $\bar{B}$ is non-compact.
In these two cases the motion of a particle is oscillating.

In \cite{IM}  we proposed the simple geometric criterion  for
finiteness of  volume of $B$ and compactness of $\bar{B}$
in terms of the positions of the
points (\ref{4.6})  with respect to
the ($N-2$)-dimensional unit sphere $S^{N-2}$
($N \geq 3$).

{\bf Definitions.}
{\em A point $\vec{n} \in S^{k}$ ($k \geq 1$) is
illuminated by a point-like source located at a point
$\vec{v} \in \R^{k+1}$, $|\vec{v}| > 1$,
if and only if $|\vec{n} - \vec{v}| \leq \sqrt{|\vec{v}|^2 -1}$.
A point $\vec{n} \in S^{k}$ is
strongly illuminated by a point-like source located at the point
$\vec{v} \in \R^{k+1}$, $|\vec{v}| > 1$, if and only if $|\vec{n} -
\vec{v}| < \sqrt{|\vec{v}|^2 -1}$.  The subset $P \subset S^{k}$  is
called (strongly) illuminated by a point-like sources at $\{
\vec{v}^{\alpha}, \alpha \in A \}$  if and only if any point from $P$ is
(strongly) illuminated by some source at $\vec{v}^{\alpha}$, $\alpha \in A
$. }

{\bf Proposition \cite{IM}.}
{\em The billiard $B$ (\ref{4.4}) has a finite
volume if and only if  point-like
sources of light located at the points $\vec{v}^{\alpha}$ (\ref{4.6})
illuminate the unit sphere $S^{N-2}$.
The closure of a billiard  $\bar{B}$ is compact
(in the topology of $D^{N-1} \simeq H^{N-1}$) if and only if
sources at points (\ref{4.6}) strongly illuminate $S^{N-2}$.}

We remind, that the problem of illumination of a convex body in a
multidimensional vector space by point-like sources for the first time was
considered in \cite{Sol,BG}. For the case of $S^{N-2}$ this problem is
equivalent to the problem of covering  spheres with spheres
\cite{Fe,R}.

There exists a topological bound on a number of point-like
sources $m$  illuminating the sphere $S^{N-2}$
\cite{BG}:
\beq{3.40o}
m \geq N.
\eeq
Thus, we obtain the restriction  (\ref{1.5}).
According to this restriction the number of $p$-branes $m =|S|$
should at least exceed the critical value $N = n+l$ for
the existence of oscillating (e.g. stochastic) behaviour
near the singularity.

We remind that Kasner-like solutions
have the following form
\bear{4.8}
&&g = w d\tau \otimes d\tau + \sum_{i=1}^{n} A_i \tau^{2 \alpha^i} g^i,
\\   \label{4.9}
&&\varphi^{\beta} =  \alpha^{\beta} \ln \tau + \varphi^{\beta}_0,
\\   \label{4.10}
&& \sum_{i=1}^{n} d_i \alpha^i =
\sum_{i=1}^{n} d_i (\alpha^i)^2 +
\alpha^{\beta} \alpha^{\gamma} h_{\beta \gamma}= 1,
\\   \label{4.11}
&& F^a= 0
\ear
where $A_i > 0$,  $\varphi^{\beta}_0$ are constants
$i = 1, \ldots, n$; $\beta, \gamma = 1, \ldots, l$;
$a \in \Delta$.  These  solutions correspond to zero
$p$-brane charges.
If the  vector of Kasner parameters $\alpha = (\alpha^{A}) =
(\alpha^{i}, \alpha^{\gamma})$ obeys the relations
\beq{4.12}
U^s(\alpha) =  U_A^{s} \alpha^A = \sum_{i\in I_s}d_i\alpha^i
 -\chi_s\lambda_{a_s \gamma}\alpha^{\gamma} > 0,
\eeq
$s \in S$, then the field configuration (\ref{4.8})-(\ref{4.11})
is the  asymptotical (attractor) solution for a family of (exact)
solutions with  non-zero charges: $Q_s \neq 0$,  when $\tau \to +0$.

Relations (\ref{4.12})  may be  easily
understood  using relations  (\ref{5.29n}).
Indeed, from (\ref{5.29n}) and  zero value limits for  forms
$F^a$, $a \in \Delta$, we get
\beq{4.12b}
\exp[2U^s(x)] = C_s \tau^{2U^s(\alpha)} \to 0,  \quad C_s \neq 0,
\eeq
for $\tau \to +0$.  These relations imply (\ref{4.12}).

Now we give a rigorous explanation of (\ref{4.12}).
Let us denote by  $K $  a
set of Kasner vector parameters  $\alpha =(\alpha^{A}) \in \R^{N}$
satisfying (\ref{4.10}).  $K$ is an ellipsoid isomorphic to $S^{N-2}$.
The isomorphism is defined by the relations
\beq{4.13}
\alpha^A =  e^A_a n^a / q, \quad  (n^a) = (1, \vec{n}),
\quad \vec{n} \in S^{N-2}.
\eeq
Here we use the diagonalizing matrix $(e^A_a)$ and the parameter $q$
 defined in subsection 3.1 (see (\ref{2.30o})).

{\bf Proposition 1.} {\em Let us consider a point-like source
located at a point $\vec{v}^s$ from (\ref{4.6}).
A point  $\vec{n} \in S^{N-2}$ is  illuminated
by this source if and only if the corresponding Kasner vector
$\alpha =(\alpha^{A}) \in K$  defined by (\ref{4.13})
satisfies the relation
\beq{4.14}
U^s(\alpha) =  U_A^{s} \alpha^A =
\sum_{i\in I_s}d_i \alpha^i
 -\chi_s \lambda_{a_s \gamma} \alpha^{\gamma} \leq 0.
\eeq
}

{\bf Proof.}
A point $\vec{n} \in S^{N-2}$  is  illuminated
by a point-like source of light located at a point $\vec{v}^s$,
$|\vec{v}^s| > 1$, if and only if
\beq{4.15}
\vec{n} \vec{v}^s  \geq 1,
\eeq
$s \in S$.
For  $ \vec{v}^s = - \vec{U}^s/U_{0}^s$, $U^s_{0} >0$, this inequality
may be rewritten as
\beq{4.16}
n^a U_a^s = q e_A^a \alpha^A U_a^s = q  \alpha^A U_A^s \leq 0,
\eeq
$s \in S$. Thus, we obtain the relation (\ref{4.14}).

{\bf Corollary.}
{\em A point  $\vec{n} \in S^{N-2}$ is not illuminated
by a source at a point $\vec{v}^s$ from (\ref{4.6})
if and only  the relation (\ref{4.12}) is satisfied.}

A small modification of the Proposition 1
is the following one.

{\bf Proposition 1a.}
{\em A point  $\vec{n} \in S^{N-2}$ is
strongly illuminated  by a source  at $\vec{v}^s$
if and only if $U^s(\alpha)  < 0$.}

Due to Proposition 1 the criterion of  the finiteness of
a billiard volume (see Proposition)
may be reformulated
in terms of inequalities on  Kasner-like parameters \cite{IM}.

{\bf Proposition 2.}
{\em Billiard $B$ (\ref{4.4}) has a finite volume if and only if
there are no $\alpha$  satisfying the
relations  (\ref{4.10}) and (\ref{4.12}).}

The positions of sources are defined (up to $O(N-1)$-rotation)
by  scalar products
\bear{4.8b}
\vec{v}^{s} \vec{v}^{s'} =
\frac{\vec{U}^{s} \vec{U}^{s'}}{ U^{s}_0 U^{s'}_0} = \qquad
\\ \nn
= 1 + \frac{(D-2)(D-1)}{d(I_s) d(I_{s'})} \left[
d(I_s \cap I_{s'})+\frac{d(I_s)d(I_{s'})}{2-D}
+\chi_s \chi_{s'} \lambda_{a\alpha} \lambda_{b\beta}
h^{\alpha\beta} \right].
\ear

Thus, we obtained a billiard representation
for the model under consideration when the
restrictions (\ref{4.1n}) and (\ref{4.2n}) are
imposed.

Now we relax the first restriction,
i.e. we put
\beq{4.1m}
\eps_s = -1,
\eeq
for some $s \in S$. Relation (\ref{4.1m}) occurs when
spherically symmetric solutions with p-branes are considered
\cite{IMJ,IMJ1}.  In this case we may obtain ``waterfall
potentials'' with $V = - \infty$ instead of  $V = + \infty$
inside of ``walls''. The ``waterfall potentials'' prevent
the oscillating behaviour near the singularity but meanwhile
do not forbid the existence of solutions with Kasner-like
asymptotical behaviour  (if there are open shadow zones).
Let us consider the following example:
\beq{4.1r}
1 \in I_s, \qquad d_1 =1,
\eeq
for all $s \in S$, i.e. all ``branes'' overlap the one-dimensional
space $M_1$. In this case the Kasner set
\beq{4.1o}
\alpha^1 = 1, \quad \alpha^i = \alpha^{\beta} = 0,
\eeq
$i =2, \ldots, n$; $\beta =1, \ldots, l$, satisfies
(\ref{4.10}) and (\ref{4.12}).
For any $\eps_s = \pm 1$, $s \in S$,
we get a family of solutions with asymptotical
Kasner-like behaviour with parameters $\alpha=(\alpha^A)$
belonging to an open neighbourhood of the (Milne) point
from (\ref{4.1o}).

\section{Examples of two-dimensional billiards}

In this section we give several examples of two-dimensional
billiards with finite areas
that occur in the models under consideration.

\subsection{Billiard is a  square}

Here we consider a model defined on the manifold
\beq{4.8a}
M = \R  \times M_{1} \times M_{2}
\eeq
governed by the  Lagrangian
\beq{4.9a}
{\cal L}=  {R}[g]  - 2\Lambda -
g^{MN} \partial_{M} \varphi \partial_{N} \varphi
- \frac{1}{n_1!} \exp[ 2 \lambda_{1} \varphi ] (F^1)^2_g
- \frac{1}{n_2!} \exp[ 2 \lambda_{2} \varphi ] (F^2)^2_g,
\eeq
where $d_{1} = \dim M_1 = d$, $d_{2} = \dim M_2 = d$, $D = 1+ 2d$,
$n_1 = d+1$, $n_2 = d$, $d \geq 2$,
$w =-1$ and $\eps(1) = \eps(2) = 1$.
Let
\beq{4.9b}
s_1 = (1,e,\{1 \}),\quad s_2 = (1,e, \{2 \}), \quad
s_3 = (2,m,\{1 \}), \quad s_4 = (2,m,\{2 \}),
\eeq
i.e. we have two electric branes  corresponding to the form $F^1$,
and two magnetic  branes  corresponding to $F^2$. Branes
$s_1$ and $s_3$ ``live'' in $M_1$ and  branes
$s_2$ and $s_4$ ``live'' in $M_2$.

We put
\beq{4.10b}
\lambda_{1} = \lambda_{2} = \lambda, \qquad   \lambda^2 = d/2.
\eeq
Then from (\ref{4.8b}) we get
\beq{4.10c}
(\vec{v}^{s_i})^2 = - \vec{v}^{s_1} \vec{v}^{s_4} =
- \vec{v}^{s_2} \vec{v}^{s_3} = 2 (2d-1),
\eeq
$i = 1,2,3,4$,
and
\beq{4.11a}
\vec{v}^{s_1} \vec{v}^{s_2} =
\vec{v}^{s_1} \vec{v}^{s_3} =
\vec{v}^{s_2} \vec{v}^{s_4} =
\vec{v}^{s_3} \vec{v}^{s_4} = 0.
\eeq

\begin{figure}[cht]
\begin{center}
\epsfig{file=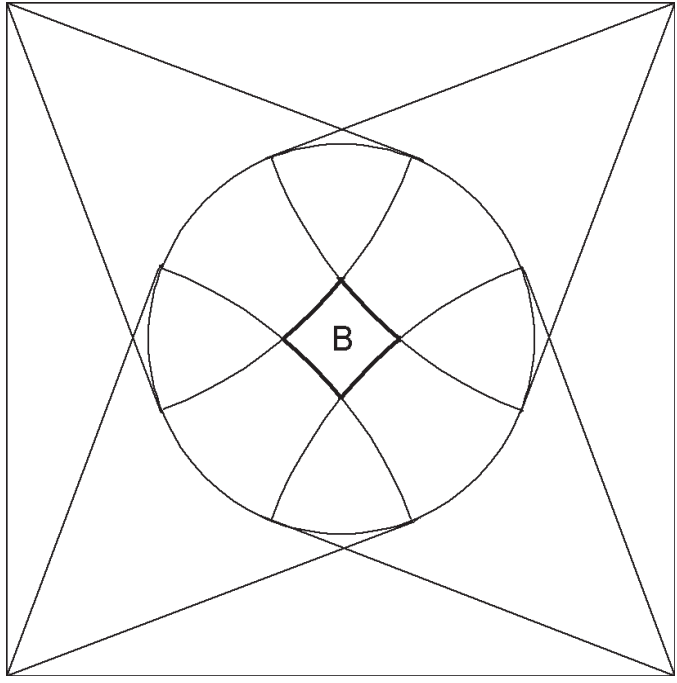,height=9cm,width=9cm}
\end{center}
\caption{
Square billiard in the 5-dimensional model with two internal spaces,
 scalar field and four ``branes'' (two electric and two magnetic).}
\end{figure}

%
%

This means that the points  $\vec{v}^{s_i}$, $i = 1,2,3,4$,
form a square in $\R^2$  containing $S^1$ ($\vec{v}^{s_1} = -
\vec{v}^{s_4}$, $\vec{v}^{s_2} = - \vec{v}^{s_3}$), i.e. all
points of $S^1$ are illuminated by these four points. The billiard
$B$ (\ref{4.4})  is a sub-compact square
($\bar{B}$ is compact) in the Lobachevsky space.
For $d = 2$ ($D=5$) it is depicted on Fig. 1.

\subsection{Billiard is a triangle}

Let us consider a model defined on
\beq{4.12a}
M = \R  \times M_{1} \times M_{2} \times M_{3}
\eeq
and governed by the Lagrangian
\beq{4.13a}
{\cal L}=  {R}[g]  - 2\Lambda
- \frac{1}{n_1!}  (F^1)^2_g ,
\eeq
where $d_{i} = \dim M_i = d$,
$n_1 = d+1$, $d \geq 2$,
$w =-1$, $\eps(i) = 1$, $i =1, 2, 3$, and $D = 1+ 3d$.
Let
\beq{4.14a}
s_1 = (1,e,\{1 \}),\quad s_2 = (1,e, \{2 \}), \quad
s_3 = (1,e,\{3 \}),
\eeq
i.e. we have three electric branes  corresponding to
the form $F^1$. The brane $s_i$  ``lives'' in $M_i$, $i =1, 2, 3$.

From (\ref{4.8b}) we get
\bear{4.15a}
(\vec{v}^{s_i})^2 =  6d -2,
\\ \label{4.16a}
\vec{v}^{s_i} \vec{v}^{s_j} = 1 - 3d,
\ear
$i \neq j$; $i,j = 1,2,3$.

For $d \geq 2$ the points  $\vec{v}^{s_i}$, $i = 1,2,3$,
form a triangle in $\R^2$  containing $S^1$ and all
points of $S^1$ are illuminated by these three points. The billiard
(\ref{4.4})  is a sub-compact triangle in the Lobachevsky space $D^2$.
For $d = 2$ it is depicted on Fig. 2.

\begin{figure}[cht]
\begin{center}
\epsfig{file=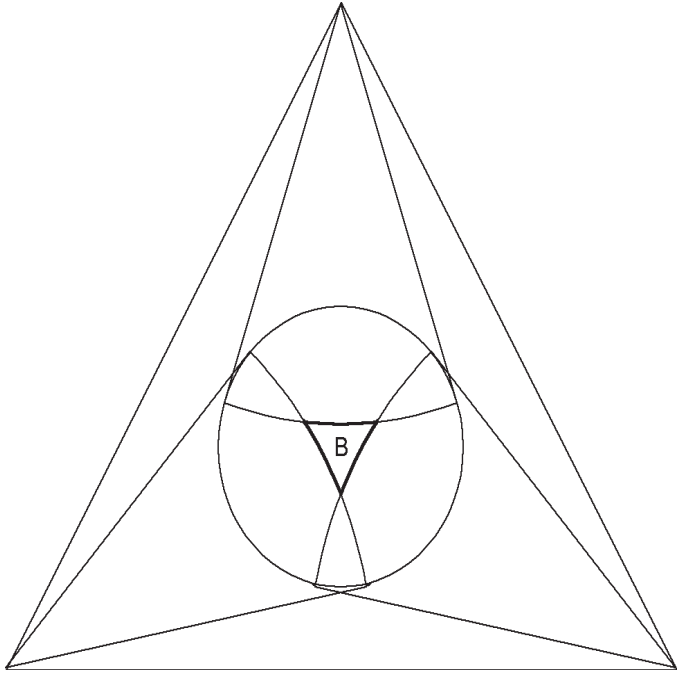,height=9cm,width=9cm}
\end{center}
\caption{
Triangle billiard in the 7-dimensional model
with three internal spaces and three electric ``branes''.}
\end{figure}

%
%

For $d = 1$ we obtain (at least formally) the billiard
$B$ depicted on Fig. 3.
The closure of $B$
is not compact but the area of $B$ is finite.
This billiard appears in the well-known Bianchi-IX model
(for a review, see \cite{P,K1,IM}).
For $d = 1$ the restriction on composite $p$-branes  (\ref{B.2})
is not satisfied but to avoid this obstacle we may consider non-composite
case when the Lagrangian (\ref{4.13a}) is replaced
by the following Lagrangian in the dimension $D =4$
with three   2-forms $F^a$, $a=1,2,3$,
\beq{4.13b}
{\cal L}=  {R}[g]  - 2\Lambda
- \sum_{a=1}^{3} \frac{1}{2!}  (F^a)^2_g ,
\eeq
and the relation (\ref{4.14a})
is replaced by its non-composite analogue:
\beq{4.14b}
s_1 = (1,e,\{1 \}),\quad s_2 = (2,e, \{2 \}), \quad
s_3 = (3,e,\{3 \}).
\eeq


\begin{figure}[cht]
\begin{center}
\epsfig{file=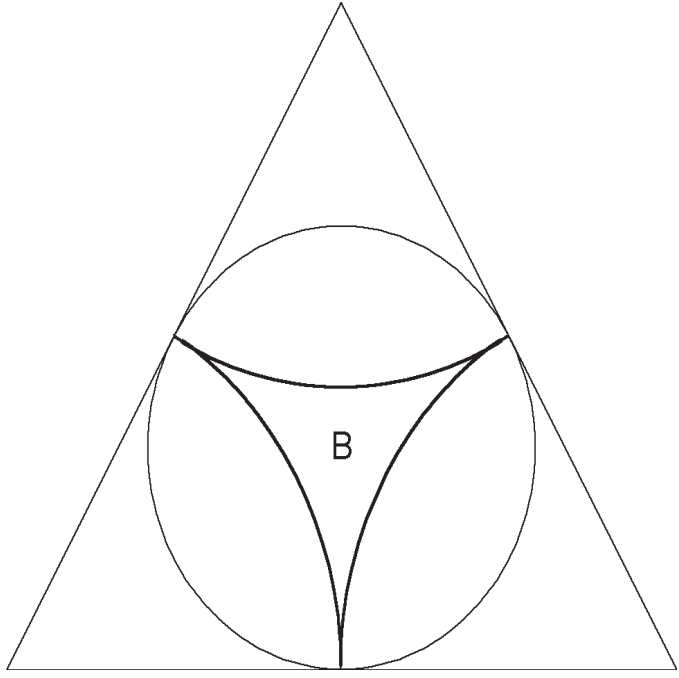,height=9cm,width=9cm}
\end{center}
\caption{
Triangle billiard for $D=4$ coinciding with that of Bianchi-IX
model.}
\end{figure}

%


\section{$D =11$ supergravity}

\subsection{``Truncated'' $D =11$ supergravity}

Now we consider a ``truncated'' bosonic sector of
$D = 11$ supergravity governed by the action
(``truncated'' means without Chern-Simons terms)
\beq{6.8}
S_{tr} =
\int d^{11}z \sqrt{|g|} \biggl\{R[g] - \frac{1}{4!} F^2_g \biggr\}
\eeq
where  $F = d A = F^1$ is a 4-form. In this case we have
electric 2-branes ($d(I_s) = 3$) and magnetic
5-branes ($d(I_s) = 6$).

From  (\ref{4.8b}) we get
\bear{6.9}
(\vec{v}^{s})^2  = 21, \qquad v_s = e,
\\ \label{6.10}
(\vec{v}^{s})^2  = 6, \qquad v_s = m,
\ear
and
\bear{6.11}
\vec{v}^{s} \vec{v}^{s'} =
&&= 1 +  10 [ d(I_s \cap I_{s'}) - 1],
\qquad   v_s = v_{s'} = e,
\\ \label{6.12}
&&= 1 +  \frac{5}{2} [ d(I_s \cap I_{s'}) - 4],
\qquad   v_s = v_{s'} = m,
\\ \label{6.13}
&&= 1 +  5 [ d(I_s \cap I_{s'}) - 2],
\qquad   v_s = e, \ v_{s'} = m.
\ear
Scalar products (\ref{6.9})-(\ref{6.13}) are (in some sense)
``building blocks'' for constructing billiards for the model
under consideration.

Now we suggest  an example of a billiard with a finite volume
that occurs in the truncated $D = 11$ supergravity.
Let us consider the metric  (\ref{2.11g})
defined on the manifold
\beq{6.14}
M = \R  \times M_{1} \times  \ldots \times M_{5},
\eeq
where all $(M_i, g^i)$, $i = 1, \ldots, 5$,
are 2-dimensional Einstein manifolds of the Euclidean signature
and $w = -1$.

We consider ten magnetic $5$-branes wrapped
on six-dimensional ``submanifolds''
$M_i \times M_j \times M_k$, $1 \leq i < j < k \leq 5$.
Thus, the  $5$-branes are  labeled by indices
\beq{6.15}
s = s(i,j,k) = (1,m, \{ i,j,k \}),
\eeq
$1 \leq i < j < k \leq 5$.

It follows from (\ref{6.10}) and  (\ref{6.12}) that
all vectors  $\vec{v}^{s}$ have the same length:
$|\vec{v}^{s}| = \sqrt{6}$ and $\vec{v}^{s} \vec{v}^{s'} = 1, -4$
for  dimensions of intersections  $d(I_s \cap I_{s'}) = 4, 2$
respectively.

Now we prove that  in our case  the set of sources located at points
$\vec{v}^{s} \in \R^4$, $s \in  S = S_m$
``illuminates'' the 3-dimensional (Kasner) sphere $S^3$
and hence (according to the Proposition from Section 4)
the 4-dimensional billiard
(\ref{4.4}) $B  \subset D^{4}$ has  a finite volume.
Due to Proposition 2 from Section 4 it is sufficient to
prove the following proposition.

{\bf Proposition 3.}
{\em There are no $\alpha = (\alpha^1, \ldots, \alpha^5) \in
\R^5$ satisfying the relations
\beq{6.16}
\sum_{i=1}^{5}  \alpha^i =
\sum_{i=1}^{5}  (\alpha^i)^2 = \frac{1}{2},
\eeq
and
\beq{6.17}
\alpha^i + \alpha^j + \alpha^k > 0,
\eeq
for all $1 \leq i < j < k \leq 5$.}

This proposition is a consequence of the following statement.

{\bf Proposition 4.}
{\em Let $\alpha = (\alpha^1, \ldots, \alpha^5) \in
\R^5$ satisfy the relations (\ref{6.16}) and
$\alpha_1 \leq \alpha_2 \leq \alpha_3 \leq \alpha_4 \leq \alpha_5$.
Then
\beq{6.18}
\alpha^1 + \alpha^2 + \alpha^3 \leq 0,
\eeq
and  $\alpha^1 + \alpha^2 + \alpha^3 = 0$ only if
$\alpha = \alpha_1$, where
\beq{6.18a}
\alpha_1 = \left(-\frac{1}{2}, \frac{1}{4},
\frac{1}{4}, \frac{1}{4}, \frac{1}{4} \right).
\eeq
}

{\bf Proof.} Let us consider the set $K$ of
Kasner vector parameters
$\alpha = (\alpha^1, \ldots, \alpha^5) \in \R^5$
satisfying (\ref{6.16}). Let
\beq{6.19}
G = \{ \alpha \in K| \alpha^1 < \alpha^2 <
\alpha^3 < \alpha^4 < \alpha^5 \}
\eeq
and
\beq{6.20}
\bar{G} = \{ \alpha \in K| \alpha^1 \leq \alpha^2
\leq \alpha^3 \leq  \alpha^4 \leq \alpha^5 \}.
\eeq
$G$ is an open submanifold of the 3-dimensional ``Kasner''
manifold $K$ and $\bar{G}$ is a closure of $G$.
$\bar{G}$ is compact subset of $K$.  Let us
consider a smooth function $f : \R^5 \rightarrow \R$ defined by
the relation
\beq{6.21}
f(\alpha) = \alpha^1 + \alpha^2 + \alpha^3.
\eeq
Let $f_{|} =f_{| \bar{G}}$ be a restriction of $f$ on $\bar{G}$.
$f_{|}$ is a continuous function reaching an (absolute) maximum
on the compact (topological) space $\bar{G}$:
\beq{6.22}
        C = {\rm max} f_{|} = f(\alpha_{max}),
\eeq
$\alpha_{max} \in \bar{G}$.  It is clear that $C \geq 0$, since
$C \geq f(\alpha_1) =0$,
where $\alpha_1$ is defined in (\ref{6.18a}).
To prove the proposition it is sufficient to prove that the point
of maximum $\alpha_{max}$  is unique and
\beq{6.23}
        \alpha_{max} =  \alpha_1.
\eeq
Let us prove the relation (\ref{6.23}).
The point $\alpha_{max}$
does not belong to $G$. Indeed, if we suppose that
$\alpha_{max} \in G$ we get the conditional extremum relation
\beq{6.24}
      d \Phi (\alpha) =  0
\eeq
at $\alpha  = \alpha_{max} $, where
\beq{6.25}
\Phi(\alpha) =  \Phi_0(\alpha) \equiv
f(\alpha) + \lambda_1 \sum_{i=1}^{5}  \alpha^i +
\lambda_2 \sum_{i=1}^{5}  (\alpha^i)^2,
\eeq
and  $\lambda_1$ and $\lambda_2$ are Langrange multipliers.
It follows from (\ref{6.24}) and  (\ref{6.25}) that
\bear{6.26}
\p_i \Phi =  1 + \lambda_1  + 2 \lambda_2 \alpha^i = 0,  \quad i= 1,2,3,
\\ \label{6.27}
\p_j \Phi =  \lambda_1  + 2 \lambda_2 \alpha^j = 0,  \quad j= 4,5,
\ear
at $\alpha  = \alpha_{max} $. Relations  (\ref{6.26}) and  (\ref{6.27})
imply $\lambda_2 \neq 0$ and  $\alpha^4 = \alpha^5$. The latter
contradicts the inequality $\alpha^4 < \alpha^5$ for points
in $G$. Thus, $\alpha_{max} \in \bar{G} \setminus G$.
The set $\p G = \bar{G} \setminus G$ is a border of
the curved tetrahedron $\bar{G}$.
It is a union of four faces
\bear{6.28}
\alpha^1=\alpha^2 < \alpha^3 < \alpha^4< \alpha^5 \quad (\Gamma_1),
\\ \label{6.29}
\alpha^1< \alpha^2 = \alpha^3 < \alpha^4< \alpha^5 \quad (\Gamma_2),
\\ \label{6.30}
\alpha^1< \alpha^2 < \alpha^3 = \alpha^4< \alpha^5 \quad (\Gamma_3),
\\ \label{6.31}
\alpha^1< \alpha^2 < \alpha^3 < \alpha^4 = \alpha^5 \quad (\Gamma_4),
\ear
six edges
\bear{6.32}
\alpha^1=\alpha^2 = \alpha^3 < \alpha^4< \alpha^5 \quad (E_{12}),
\\ \label{6.33}
\alpha^1= \alpha^2 < \alpha^3 = \alpha^4< \alpha^5 \quad (E_{13}),
\\ \label{6.34}
\alpha^1= \alpha^2 < \alpha^3 < \alpha^4= \alpha^5 \quad (E_{14}),
\\ \label{6.35}
\alpha^1<\alpha^2 = \alpha^3 = \alpha^4< \alpha^5 \quad (E_{23}),
\\ \label{6.36}
\alpha^1< \alpha^2 = \alpha^3 < \alpha^4 = \alpha^5 \quad (E_{24}),
\\ \label{6.37}
\alpha^1< \alpha^2 < \alpha^3 = \alpha^4= \alpha^5 \quad (E_{34}),
\ear
and four vertices
\bear{6.38}
\alpha^1 < \alpha^2 = \alpha^3 = \alpha^4= \alpha^5 \quad (V_1),
\\ \label{6.39}
\alpha^1 = \alpha^2 < \alpha^3 = \alpha^4= \alpha^5 \quad (V_2),
\\ \label{6.40}
\alpha^1 = \alpha^2 = \alpha^3 < \alpha^4= \alpha^5 \quad (V_3),
\\ \label{6.41}
\alpha^1 = \alpha^2 = \alpha^3 = \alpha^4< \alpha^5 \quad (V_4).
\ear
The point of maximum $\alpha_{max}$ does not belong to any
face $\Gamma_a$, $a=1,2,3,4$. Indeed, if we suppose that
$\alpha_{max}$ belongs to some  face $\Gamma_a$
we get the conditional extremum relation (\ref{6.24})
with modified $\Phi$:
\beq{6.42}
\Phi =  \Phi_0  + \lambda_3   (\alpha^a - \alpha^{a+1}),
\eeq
$a=1,2,3,4$. But one may verify that
there are no solutions of the relation (\ref{6.24}) in this case.
Analogous arguments lead us to a non-existence of
points of maximum among the points of edges
$E_{12}, E_{13}, E_{14}, E_{23}, E_{24}, E_{34}$. Here the only
difference is that there exist (only) two extremal points
belonging to $E_{13}$ and $E_{23}$  respectively:
\bear{6.33a}
\alpha^1= \alpha^2 = \frac{1}{10} \left(1 - \frac{12}{\sqrt{14}} \right),
\alpha^3 = \alpha^4 =
\frac{1}{10} \left(1 + \frac{3}{\sqrt{14}} \right),
\alpha^5 =
\frac{1}{10} \left(1 + \frac{18}{\sqrt{14}} \right),
\\ \label{6.35a}
\alpha^1
= \frac{1}{10} \left(1 - \frac{3}{2} \sqrt{6} \right),
\alpha^2 = \alpha^3 = \alpha^4 =
\frac{1}{10} \left(1 - \frac{1}{4}\sqrt{6} \right),
\alpha^5 =
\frac{1}{10} \left(1 + \frac{9}{4}\sqrt{6} \right),
\ear
but
$f(\alpha) < 0$ in these points and hence they are not
the points of maximum. Thus, $\alpha_{max}$ belongs to the
set of vertices:
\bear{6.38a}
\alpha^1 = - \frac{1}{2}, \quad
\alpha^2 = \alpha^3 = \alpha^4= \alpha^5  = \frac{1}{4} \quad (V_1),
\\ \label{6.39a}
\alpha^1 = \alpha^2 =
\frac{1}{10} \left(1 - \frac{3}{2} \sqrt{6} \right), \quad
\alpha^3 = \alpha^4= \alpha^5 =
\frac{1}{10} \left(1 + \sqrt{6} \right) \quad (V_2),
\\ \label{6.40a}
\alpha^1 = \alpha^2 = \alpha^3 =
\frac{1}{10} \left(1 - \sqrt{6} \right), \quad
\alpha^4= \alpha^5 =
\frac{1}{10} \left(1 + \frac{3}{2} \sqrt{6} \right) \quad (V_3),
\\ \label{6.41a}
\alpha^1 = \alpha^2 = \alpha^3 = \alpha^4 = - \frac{1}{20},  \quad
\alpha^5 = \frac{7}{10} \quad  (V_4).
\ear
The calculation gives us  $f(\alpha) < 0$  for
$\alpha = V_i$, $i =2,3,4$. Thus, the first vertex $V_1 = \alpha_1$
is the point of maximum. The Proposition 4 is proved.

We also  proved the Proposition 3  (as a consequence
of the Proposition 4) and
due to Proposition 2 the following proposition.

{\bf Proposition 5.} {\em  For the model (\ref{6.8})-(\ref{6.15})
the Kasner sphere $S^3$ is illuminated
by the set of ten sources located at points
$\vec{v}^{s} \in \R^4$ from (\ref{4.6}) , with $s = s(i,j,k)$,
$1 \leq i < j < k \leq 5$, defined in (\ref{6.15}),
and hence the billiard $B$ (\ref{4.4}) has a finite volume.}

Moreover, this proposition may be strengthen as follows.

{\bf Proposition 5a.} {\em In terms of  Proposition 5
all points of the Kasner $S^3$ sphere
except   five points $\vec{n}_1, \ldots, \vec{n}_5
\in S^3$ corresponding to  the Kasner sets
\beq{6.18b}
\alpha_1 =
\left(-\frac{1}{2}, \frac{1}{4}, \frac{1}{4}, \frac{1}{4}, \frac{1}{4}
\right), \ldots, \alpha_5 = \left( \frac{1}{4}, \frac{1}{4}, \frac{1}{4},
\frac{1}{4}, -\frac{1}{2} \right) \in K,
\eeq
respectively
(see (\ref{4.13})) are strongly illuminated by  the set of ten sources
$\vec{v}^{s} \in \R^4$, $s \in S$.}

{\bf Proof.} The set $K$ of Kasner parameters
satisfying  (\ref{6.16}) is a union of $5! = 120$ ``sectors''
\beq{6.18c}
K = K_{12345} \cup K_{21345} \cup \ldots \cup K_{54321},
\eeq
where
\beq{6.18d}
K_{i_1 i_2 i_3 i_4 i_5} =
\bar{G} = \{ \alpha \in K| \alpha^{i_1} \leq \alpha^{i_2}
\leq \alpha^{i_3} \leq  \alpha^{i_4} \leq \alpha^{i_5} \}
\eeq
and $(i_1, i_2, i_3, i_4, i_5)$ is a permutation of $(1, 2, 3, 4, 5)$.
Due to  Proposition 1a and  Proposition 4
any point $\vec{n} = \vec{n}(\alpha) \in S^3$ corresponding to
$\alpha \in K_{12345} \setminus \{ \alpha_1 \}$  is strongly
illuminated by the source located at point
$\vec{v}^{s}$,  with $s = s(1,2,3)$.
Analogously, any point $\vec{n} =
\vec{n}(\alpha) \in S^3$ corresponding to
$\alpha \in K_{i_1 i_2 i_3 i_4 i_5} \setminus \{ \alpha_{i_1} \}$  is
strongly illuminated by the source located at point $\vec{v}^{s}$,  with
$s = s(i_1,i_2,i_3)$, where
$(i_1, i_2, i_3, i_4, i_5)$ is a permutation of $(1, 2, 3, 4, 5)$.
The proposition is proved.

The points $\vec{n}_1, \ldots, \vec{n}_5$ from Proposition 5a
are not strongly illuminated (see Proposition 1a). Thus,
Proposition 5,  Proposition 5a and  Proposition (from Section 4)
imply that the billiard $B$ (\ref{4.4}) has a finite volume
but its closure $\bar{B}$ is not compact.
Points $\vec{n}_1, \ldots, \vec{n}_5 \in S^3$ are
ending points of five ``horns'' of $B$.
(These ``horns''  look similar to potential energy
``valleys'' that occur in some toy models, e.g.
related to M(atrix) theory \cite{AMRV}).

Using relations (\ref{4.13}) one
may verify that
\beq{6.8b}
\vec{n}_i \vec{n}_j  = - \frac{1}{4}, \quad i \neq j,
\eeq
$i,j = 1, \ldots, 5$. This means that
$\vec{n}_1, \ldots, \vec{n}_5$ are vertices
of a 4-dimensional simplex.

Here we considered the billiard $B$
generated by ten sources, corresponding to
non-zero charges :
$Q_s \neq 0$, $s \in  S$. If some charges are  zero:
$Q_s = 0$, $s \in  S_0$,
then the corresponding points $\vec{v}^{s}$,
$s \in  S_0$ ($S_0 \neq \emptyset$)
are ``switched off'' and billiard is
generated by $m \leq 9$ point-like sources. In this
case we have the following proposition.

{\bf Proposition 5b.}  {\em In terms of  Proposition 5
any subset of sources $\vec{v}^s$, $s \in  S \setminus S_0$,
with $S_0 \neq \emptyset$, does not illuminate
the Kasner  sphere $S^3$  and hence the billiard $B$
generated by this subset has an infinite volume.}

{\bf Proof.}  Without a loss of generality
let us suppose that $S_0$ contains $s(1,2,3)$,
i.e.  at least the source at $\vec{v}^s$, $s=s(1,2,3)$,
is ``switched off''. Then, the point $\vec{n}$ corresponding to
the  set $\alpha = (\alpha^i)$ from
(\ref{6.40a}) belongs to the shadow, since
\beq{6.17a}
\alpha^i + \alpha^j + \alpha^k > 0,
\eeq
for all $1 \leq i < j < k \leq 5$, $(i,j,k) \neq (1,2,3)$.
Thus, the shadow set is non-empty. (This set
is open since it is an intersection
of a finite number of open shadow sets corresponding to sources of light.)
The proposition is proved.

\subsection{Inclusion of Chern-Simons term}

Now we consider the total bosonic sector action
for  $D =11$ supergravity with the Chern-Simons term included:
\beq{6.8a}
S =  S_{tr} +  c \int_{M} A \wedge F \wedge F
\eeq
where $S_{tr}$ is defined in (\ref{6.8}),
$c = {\rm const}$,  ($F = d A$).
Since the second term in (\ref{6.8a}) (Chern-Simons term)
does  not depend upon a metric the Einstein equations
are not changed. The only modification  of  equations  of motion
is related to ``Maxwell'' equations
\beq{6.43}
d*F = {\rm const} \ F \wedge F.
\eeq
Due to (\ref{6.43})   solutions
to field equations corresponding to the truncated
model  (\ref{6.8})  with the trivial Chern-Simons term
\beq{6.44}
F \wedge F = 0.
\eeq
are also solutions for $D =11$ supergravity.

Now, we are interested in existence of solutions
for the truncated model with non-zero charges $Q_s \neq 0$
satisfying (\ref{6.44}).

Calculating the Hodge dual in (\ref{2.29n})
and using the solution (\ref{5.29n})  we get
\beq{6.45}
F = - e^f \sum_{s \in S} Q_s \tau(\bar{I}_s), \quad
f = 2 \sum_{i=1}^{5}x^i - \gamma,
\eeq
where $\bar{I}_s = \{1,2,3,4,5 \} \setminus I_s$, $s \in S =S_m$,
and $S$ is the index set defined in (\ref{6.15}).
For the Chern-Simons term we obtain
\beq{6.46}
e ^{-2f} F \wedge F = 2 \sum_{i = 1}^{5} P_i \tau(\bar{\{i \}}),
\eeq
where
\beq{6.46b}
2P_i = \sum_{s,s' \in S} Q_s Q_{s'}   \delta(I_s \cap I_{s'}, \{i \}),
\eeq
$i =1,2,3,4,5$.  Here $\delta(A,B) =1$ for $A = B$,
$\delta(A,B) = 0$ otherwise.

{\bf Proposition 6.} {\em Let all charges  be non-zero:
$Q_s \neq 0$ for all $s \in S$. Then
\beq{6.44b}
(F \wedge F)(z) \neq 0,
\eeq
in all points $z \in M$.}

{\bf Proof.} Let us suppose that the Chern-Simons term vanishes
in some point (i.e. relation (\ref{6.44}) in some point is satisfied).
Since forms $\tau(\bar{\{i \}})$
are linearly independent at any point,
we obtain $P_i = 0$, $i =1,2,3,4,5$,
or explicitly
\bear{6.45a}
P_1 = Q_{123} Q_{145} +  Q_{124}Q_{135}+ Q_{125}Q_{134} = 0,
\\ \label{6.46a}
P_2 = Q_{123} Q_{245} +  Q_{124}Q_{235}+ Q_{125}Q_{234} = 0,
\\ \label{6.47a}
P_3 = Q_{123} Q_{345} +  Q_{134}Q_{235}+ Q_{135}Q_{234} = 0,
\\ \label{6.48a}
P_4 = Q_{124} Q_{345} +  Q_{134}Q_{245}+ Q_{145}Q_{234} = 0,
\\ \label{6.49a}
P_5 = Q_{125} Q_{345} +  Q_{135}Q_{245}+ Q_{145}Q_{235} = 0,
\ear
where we denote $Q_{ijk} = Q_s$ for $s = s(i,j,k)$,
$1 \leq i < j < k \leq 5$. Let us denote
$k_2 = Q_{345}$, $k_3 = Q_{245}$, $k_4 = Q_{235}$, $k_5 = Q_{234}$
and $a = Q_{145}$,  $b = Q_{135}$. From (\ref{6.49a}) we
get
\beq{6.50}
Q_{125} = - \frac{k_3}{k_2} b -  \frac{k_4}{k_2} a.
\eeq
From the relation $k_3 P_3 +k_4 P_4 +k_5 P_5 = 0$ we get
\beq{6.51}
Q_{134} = - \frac{k_5}{k_4} b -  \frac{k_5}{k_3} a.
\eeq
From (\ref{6.47a})and (\ref{6.48a}) we
deduce
\beq{6.52}
Q_{123} = \frac{k_4 k_5}{k_2 k_3} a, \quad
Q_{124} = \frac{k_3 k_5}{k_2 k_4} b.
\eeq
Substituting  (\ref{6.50})-(\ref{6.52}) into eq. (\ref{6.45a})
we get
\beq{6.53}
(k_4 a)^2 + (k_3 b)^2 +  (k_4 a + k_3 b)^2 = 0.
\eeq
Hence $a=b =0$. But this contradicts our supposition, that
$Q_s \neq 0$. So, the proposition is proved.

Thus, it follows from   Proposition 5b and Proposition 6,
that the inclusion of the Chern-Simons term leads us to
the billiard $B$ of infinite volume. In this case some
Kasner (shadow) zones are opened and we have the Kasner-like
behaviour near the singularity.

\section{Discussions}

In this paper we considered the behaviour near the
singularity of  the multidimensional model describing the cosmological
evolution of  several Einstein spaces in the theory with  scalar fields
and fields of forms.  Using the results from \cite{IKM1,IKM,IM} we
obtained the billiard representation on  multidimensional Lobachevsky
space for the cosmological model near the singularity.  We suggested
and studied
examples with oscillating behaviour near the singularity in the model with
four $p$-branes and a  square billiard and in the model with three
$p$-branes, when the billiard is a triangle (like it takes place in
Bianchi-IX model). A 4-dimensional billiard with a finite volume in the
``truncated'' $D = 11$ supergravity was also considered. It was shown
that the inclusion of the Chern-Simons term leads to the destruction
of some  confining walls of the billiard.

\bigskip

\noi
{\bf Acknowledgments.}

This work was supported in part
by the Russian Ministry for
Science and Technology and  Russian Fund for Basic Research,
project N 98-02-16414.  V.N.M. is grateful
to the Dept. Math., University of Aegean for the kind
hospitality during his stay in Karlovassi, Greece in December
1998 - January 1999.

\end{document}